\documentclass[twocolumn,showpacs,preprintnumbers,amsmath,amssymb,superscriptaddress,graphicx,graphics,pstricks]{revtex4}
\usepackage{amsmath}
\usepackage{graphicx}
\usepackage[latin1]{inputenc}
\usepackage{multido}
\usepackage{dcolumn}
\usepackage{bm}
\usepackage{epsfig}

\begin{document}

\title{The influence of a surface in the non-retarded interaction between two atoms}

\title{The influence of a surface in the non-retarded interaction between two atoms}

\author{Reinaldo de Melo e Souza}
\author{W.J.M. Kort-Kamp}
\author{F.S.S. Rosa}
\author{C. Farina}
\affiliation{Instituto de Fisica, UFRJ, CP 68528, Rio de Janeiro,
RJ, 21941-972, Brasil}

\begin{abstract}
In this work we obtain analytical expressions for the non-additivity effects in the dispersive
interaction between two atoms and perfectly conducting
surface of arbitrary shape in the non-retarded regime. We show that this three
bodies quantum-mechanical problem can be solved by mapping it into a two-bodies electrostatic one. We apply the general
formulas developed in this paper in several examples. Firstly we re-derive the London interaction as a particular
case of our formalism. Then we treat two atoms in the presence of a plane, re-obtained the result displayed in the
literature. After we add some new examples. A particularly interesting one is two atoms inside a plate capacitor,
a situation where non-additivity is very manifest since the plates lead to the
exponentially suppression of the interaction of the atoms, provided the atoms
are separated by a distance of the order of the distance between the plates or greater. 
Our results holds even in the presence of other atoms inside the plate capacitor.
As a last example we deal with two atoms in the presence of a sphere, both grounded and isolated. We show
that for realistic experimental parameters the non-additivity may be relevant for the force in each atom.
 \end{abstract}


\maketitle
%
%
%
%
\section{Introduction}

One of the most interesting features of the van der Waals dispersive
 interactions is their non-additivity\cite{israel}-\cite{mahanty}, pointed out for the first time in\cite{Axilrod}
 (for a pedagogical exposition see \cite{ajp}).
It means that the interaction between three bodies doesn't follow from the superposition principle,
or, equivalently, that the presence of a body influences the interaction of others. Although such effects have been known for many
 decades, their consequences are not yet fully investigated and there are still a lot of research trying to figure them 
 out. One of the fertile soils that is being explored is the so-called Efimov quantum state\cite{efimov}, in which
 a resonant two-body force between identical bosons can produce bound states in a three-body system even if there isn't any corresponding two-body bound state. The utilization of Feshbach resonances to tune the interaction in ultracold atom systems has allowed to experimentally probe the Efimov states\cite{Kraemer}-\cite{Huang}. These are part of a larger programme which studies universal properties of few-body systems with large scattering length,\cite{braaten}-\cite{kievsky}. It is also well-known that the interaction between two Rydberg atoms can inhibit all but a single collective
 Rydberg excitation, a phenomenon called dipole blockade \cite{lukin}-\cite{comparat}. It has recently been shown
 that the inclusion of a third Rydberg atom can break the dipole blockade\cite{pohl} due to non-additivity effects.
Another very interesting consequence of non-additivity is the great enhancement 
of the dispersive force between atoms in the vicinities of a one-dimensional transmission line\cite{f}.

In the literature there are
numerous papers dedicated to study the non-additivity\cite{mcbla}-\cite{Milton2}
in some simple systems. 
The particular case of the three-atom problem is presented in many textbooks\cite{margeneau}-\cite{mahanty}. 
Unfortunately, due to major calculation difficulties, 
there are not many cases analysed, specially those involving macroscopic bodies. 
A method developed by C. Eberlein and R. Zietal\cite{Eberlein2007} enables us
to evaluate the non-retarded dispersive interaction between one atom and a perfectly conducting surface of arbitrary shape, requiring
only the knowledge of a classical Green function that can be obtained from an electrostatic problem. This method
has been applied in a variety of interesting problems\cite{Eberlein2009}-\cite{nos}. In this
paper we generalize this method in order to have two atoms interacting with a surface, and thus
we are able to obtain analytically the influence exerted by a given
perfectly conducting surface on the interaction of two atoms in the non-retarded regime.
As a particular case, we show that in the absence of surfaces we recover the well-known
interaction between two-atoms, namely, London's formula. 
This paper is organized as follows. In the next section we generalize Eberlein-Zietal's method
to include a second atom in the system. We then identify
the general expression for the non-additivity term and show how it is related to an electrostatic problem
of a {\it single} charge in the presence of the conducting body. In section III we evaluate the influence of
a surface on the interatomic interaction and in the subsequent section we analyse some examples. Firstly we
re-obtain the interaction energy for two atoms in the presence of an infinite conducting plane. Then we analyse two
atoms inside two parallel infinite planes. This example is relevant experimentally\cite{hinds} and we show that
the non-additivity effects are strongly perceivable, leading to an exponentially suppression of the
van der Waals force between the atoms, provided 
they are kept apart by distances of the order of the separation between the
plates or larger. This has far reaching consequences. For instance, the van der Waals equation
for gases presents deviations from ideal gases by taking into account the finite volume of atoms and the
interatomic interactions. Hence, our results show that by placing a gas inside a plate capacitor it behaves
more ideally. A similar exponential attenuation  
was also obtained in\cite{f2} in the retarded regime,
for atoms inside a rectangular box. As a last example, we display the calculation
of two atoms in the presence of a conducting sphere, both grounded and isolated.
We leave a final section for conclusions and final remarks.

\section{Interaction energy for two atoms and a conducting surface}

	
In the non-retarded regime the electromagnetic field doesn't have to be quantized. Therefore the interaction
of an atom and a surface, which is usually dealt within a quantum electrodynamics framework, can
instead be approached by standard quantum mechanics techniques, where the interaction hamiltonian to be used 
in perturbation calculations is given by the instantaneous Coulomb interaction\cite{cohen}.
The convenience of Eberlein-Zietal
procedure is to relate the quantum mechanical problem to an electrostatic one, allowing us to solve
the non-retarded interaction in the simpler electrostatic domain. 

To begin with, let us consider two atoms $A$ and $B$ at positions $\mathbf{r}_A$ and
$\mathbf{r}_B$, respectively, in the presence of a grounded surface $\mathcal{S}$. 
The electrostatic energy of the configuration is given by
\begin{equation}
	U = \frac{1}{2}\int \rho({\bf r}) \Phi({\bf r})\, d^3{\bf r}\, , \label{u}
\end{equation}
where $\rho({\bf r})$ gives the charge distribution and $\Phi(\mathbf{r})$ is the electrostatic potential, which satisfies Poisson equation
\begin{equation}
	\nabla^2\Phi(\mathbf{r}) = -\frac{\rho(\mathbf{r})}{\varepsilon_0} \label{Poisson}
\end{equation} 
and vanishes at surface $\mathcal{S}$.
The solution of equation (\ref{Poisson}) can be written
in terms of the Green function 
\begin{equation}\label{phig}
\Phi(\mathbf{r}) = \frac{1}{\varepsilon_0}\int G(\mathbf{r},\mathbf{r}') \rho (\mathbf{r}') d^3\mathbf{r}' 
\end{equation}
where $G(\mathbf{r},\mathbf{r}')$ is the solution of the differential equation
\begin{equation}
	\nabla^2G(\mathbf{r},\mathbf{r}') = -\delta(\mathbf{r}-\mathbf{r}') \, , \label{green}
\end{equation}
subjected to the boundary
condition
\begin{equation}\label{ccgat}
G(\mathbf{r},\mathbf{r}')\Bigg|_{\mathbf{r}\in\mathcal{S}}=0 \, .
\end{equation}
By substituting eq.(\ref{phig}) into eq.(\ref{u}), we write the electrostatic energy as 
\begin{equation}\label{energyg}
U = 
\frac{1}{2\varepsilon_0}\int\!\! d^3\mathbf{r}\, d^3\mathbf{r}'\;\rho(\mathbf{r}) G(\mathbf{r},\mathbf{r}')\rho(\mathbf{r}')  \, .
\end{equation}
An immediate particular solution of the equation (\ref{green}) is $1/4\pi|\mathbf{r}-\mathbf{r}'|$ which, however, doesn't obey
the boundary condition (\ref{ccgat}). This suggests that we decompose our Green function in the form
\begin{equation}\label{G}
G(\mathbf{r},\mathbf{r}')=\frac{1}{4\pi|\mathbf{r}-\mathbf{r}'|}+G_H(\mathbf{r},\mathbf{r}')\, ,
\end{equation}
where $G_H$ satisfies the Laplace equation $\nabla^2G_H=0$ with the boundary condition
\begin{equation}\label{cch}
\left[\frac{1}{4\pi|\mathbf{r}-\mathbf{r}'|} + G_H(\mathbf{r},\mathbf{r}')\right]_{\mathbf{r}\in S}=0 \, . 
\end{equation}
The equations obeyed by $G_H$ are 
analogous to those satisfied by the potential $\Phi_i(\mathbf{r})$ generated by the image charges in the 
electrostatic problem of a charge $q$ at position $\mathbf{r}'$ in the presence of a perfectly conducting surface $S$. If we solve this electrostatic problem we will get $G_H$ from
the relation
\begin{equation}
		G_H(\mathbf{r},\mathbf{r}')=\frac{\varepsilon_0\phi_i(\mathbf{r})}{q} \, . \label{ghimages}
\end{equation}
The variable $\mathbf{r}'$ is implicitly present in $\phi_i(\mathbf{r})$ since the image charges depend upon
the position $\mathbf{r}'$ of the source charge.
As it will become clear, $G_H$ is the only function which must be calculated in order to evaluate
the non-additivity effects of our problem. Therefore, this method enables us to effectively replace a quantum mechanical
problem of two atoms in the presence of a conducting body by an electrostatic one of a single charge in
the presence of the conducting body. To proceed further we must specify the charge distribution $\rho(\mathbf{r})$
appearing in eq.(\ref{energyg}). We model each atom, in a first approximation, as an electric 
dipole. Hence, the charge distribution is given by
%
%
\begin{eqnarray}
	\rho (\mathbf{r}) &=& \lim\limits_{\mathbf{h}_A\rightarrow 0\atop{ q\mathbf{h}_A=\mathbf{d}_A} } q\left[\delta(\mathbf{r}-(\mathbf{r}_A+\mathbf{h}_A))-\delta(\mathbf{r}-\mathbf{r}_A)\right] \nonumber \\
&+&\lim\limits_{\mathbf{h}_B\rightarrow 0\atop{ q\mathbf{h}_B=\mathbf{d}_B} }q\left[\delta(\mathbf{r}-(\mathbf{r}_B+\mathbf{h}_B))-\delta(\mathbf{r}-\mathbf{r}_B)\right]  \nonumber \\
&=:& \rho_A (\mathbf{r})+\rho_B (\mathbf{r}) \label{denscargdoisat}
\, .
\end{eqnarray}
In the following we write the electrostatic energy of two point dipoles in the presence
of a conducting surface. This allows us to write the quantum hamiltonian interaction for
two atoms and a conducting surface by promoting $\mathbf{d}$ to a quantum operator. Substituting the decomposition (\ref{G}) and eq. (\ref{denscargdoisat}) into equation (\ref{energyg}) we obtain
\begin{eqnarray}
	U&=&\frac{1}{2\varepsilon_0}\int [\rho_A (\mathbf{r})+\rho_B (\mathbf{r})]G(\mathbf{r},\mathbf{r}')[\rho_A (\mathbf{r}')+\rho_B (\mathbf{r}')] d^3\mathbf{r}'d^3\mathbf{r} \nonumber \\ 
&=:& U_{A}+U_{B}+U_{crossed}\, ,  \label{vcoul3}
\end{eqnarray}
where
\begin{equation}
	U_{i}=\frac{1}{2\varepsilon_0}\int \rho_i (\mathbf{r})G(\mathbf{r},\mathbf{r}')\rho_i (\mathbf{r}') d^3\mathbf{r}'d^3\mathbf{r} \, , \label{ui}
\end{equation}
$i=A,B$ and 
\begin{equation}
	U_{crossed}=\frac{1}{\varepsilon_0}\int \rho_A (\mathbf{r})G(\mathbf{r},\mathbf{r}')\rho_B (\mathbf{r}') d^3\mathbf{r}'d^3\mathbf{r} \, .
\end{equation}
To obtain the last equation we used that $G(\mathbf{r},\mathbf{r}')=G(\mathbf{r}',\mathbf{r})$,
whose validity follows from Green's identity\cite{stakgold}. 
Since the Green function is the same in the case of one atom or two atoms, $U_i$ represents the 
interaction energy between point dipole $i$ and the surface $\mathcal{S}$ in the absence of the other dipole
\cite{self}. To unveil the physical meaning of $U_{AB}$ let's employ the decomposition (\ref{G})
\begin{eqnarray}
	U_{crossed}&=&\underbrace{\frac{1}{\varepsilon_0}\int \frac{\rho_A (\mathbf{r})\rho_B (\mathbf{r}')}{4\pi|\mathbf{r}-\mathbf{r}'|} d^3\mathbf{r}'d^3\mathbf{r}}_{U_{AB}} +\cr\cr
&+&\underbrace{\frac{1}{\varepsilon_0}\int \rho_A (\mathbf{r})G_H(\mathbf{r},\mathbf{r}')\rho_B (\mathbf{r}') d^3\mathbf{r}'d^3\mathbf{r}}_{U_{ABS}} \, . 
	\label{uab}
\end{eqnarray}
The first term on the right-hand-side of last equation doesn't depend on the surface. It describes the interaction between two point dipoles in vacuum. 
The last term depends conjointly on both dipoles and the surface. We will call it $U_{ABS}$. Therefore,
we write 
\begin{equation}
	 U=U_{A}+U_{B}+U_{AB}+U_{ABS} \, . \label{ucompleto}
\end{equation}
We see at once one positive aspect of this formalism. It enables us to study separately the so-called
non-additivity of dispersive forces, which is totally contained in the last term. Substituting 
eq.(\ref{denscargdoisat}) into eq. (\ref{ui}), performing a Taylor expansion and discarding divergent self-interaction terms, we obtain $U_A$ and $U_B$,
\begin{eqnarray}
	U_A&=&(\mathbf{d}_A\cdot\nabla')(\mathbf{d}_A\cdot\nabla)G_H(\mathbf{r},\mathbf{r}')\bigg|_{\mathbf{r}=\mathbf{r}'=\mathbf{r}_A} \nonumber \\
	U_B&=&(\mathbf{d}_B\cdot\nabla')(\mathbf{d}_B\cdot\nabla)G_H(\mathbf{r},\mathbf{r}')\bigg|_{\mathbf{r}=\mathbf{r}'=\mathbf{r}_B} \, . \label{uirenor}
\end{eqnarray}
In this way we re-obtain as a particular case Eberlein-Zietal formula \cite{Eberlein2007}  
for the interaction of a sole atom with a conducting surface. A similar 
treatment of the terms $U_{AB}$ and $U_{ABS}$ gives 
\begin{eqnarray}
\label{vlondon} U_{AB}&=&\frac{1}{\varepsilon_0}(\mathbf{d}_B\cdot\nabla')(\mathbf{d}_A\cdot\nabla)\frac{1}{4\pi|\mathbf{r}-\mathbf{r}'|}\bigg|_{\mathbf{r}=\mathbf{r}_A,\mathbf{r}'=\mathbf{r}_B} \nonumber \\
U_{ABS}&=&\frac{1}{\varepsilon_0}(\mathbf{d}_B\cdot\nabla')(\mathbf{d}_A\cdot\nabla)G_H(\mathbf{r},\mathbf{r}')\bigg|_{\mathbf{r}=\mathbf{r}_A,\mathbf{r}'=\mathbf{r}_B} \!\!\!\! . \label{vabs}
\end{eqnarray}
Until now,  we have been working within classical electrostatics. The passage for the quantum mechanical problem is performed
by promoting $\mathbf{d}$ to an operator in eqs.(\ref{ui}) and (\ref{vabs}), in order to obtain
the interaction hamiltonian for the quantum mechanical problem of two atoms and a conducting surface. Therefore,
the previous decomposition (\ref{ucompleto}) can be recast into the form
\begin{eqnarray}
\hat{H}_{int}=\hat{H}_A+\hat{H}_B+\hat{H}_{AB}+\hat{H}_{ABS} \, , \label{hintdois}
\end{eqnarray}
where the operators on the right hand side are obtained from $U_A$, $U_B$, $U_{AB}$ and 
$U_{ABS}$ by changing $\mathbf{d}$ by the quantum mechanical operator $\hat{\mathbf{d}}$.
Note that, as mentioned before, the interaction hamiltonian does not involve field operators since
we are in the non-retarded regime. To
obtain the interaction energy for the dispersive interaction between the atoms, assumed in the ground state, and 
the surface we 
proceed perturbatively. In first order of perturbation theory we have $E_{NR}^{(1)}:=\langle \hat{H}_{int}\rangle$,
where $\langle\cdots\rangle$ denotes the expectation value of the operator inside the brackets in the ground state 
$|0_A\, ,0_B\rangle$  of the atoms. In this order only the first two terms in eq.(\ref{hintdois}) contribute, 
since for atoms with no permanent dipole
moment we have 
\begin{equation}
	\langle 0_A\, ,0_B|\hat{d}^A_i\hat{d}^B_j|0_A\, ,0_B\rangle=\langle 0_A |\hat{d}^A_i|0_A\rangle\langle 0_B| \hat{d}^B_j|0_B\rangle=0 \, .
\end{equation}
From now on we omit the hats to denote quantum operators in order to not overburden the notation. 
Evaluating $\langle H_A\rangle$ and $\langle H_B\rangle$ we obtain
\begin{eqnarray}
	E_{NR}^{(1)}(\mathbf{r}_A,\mathbf{r}_B)&=&\frac{1}{2\varepsilon_0}\sum\limits_{m}\langle (d_m^A)^2\rangle\nabla_m\nabla_m' G_H(\mathbf{r},\mathbf{r}')\bigg|_{\mathbf{r}=\mathbf{r}'=\mathbf{r}_A}\nonumber \\
&+&\frac{1}{2\varepsilon_0}\sum\limits_{m}\langle (d_m^B)^2\rangle\nabla_m\nabla_m' G_H(\mathbf{r},\mathbf{r}')\bigg|_{\mathbf{r}=\mathbf{r}'=\mathbf{r}_B} \!\!\!\!\!\!\!\!\!\!\!\!\! , \label{eint1}
\end{eqnarray}
where we employed $\langle d_md_n\rangle =\delta_{mn}\langle d_m^2\rangle$, valid for orthonormal basis, which are
used throughout this paper.
In other words, in this approximation the atoms don't perceive each other and the 
interaction of the system is the direct superposition of the interaction between each atom and
the surface $S$. The non-additivity effects we are looking for appear only in the next order.
The second order contribution to the interaction energy is
\begin{equation}
	E_{NR}^{(2)}(\mathbf{r}_A,\mathbf{r}_B)=-\sum\limits_{I}\,\!'\,\, \frac{\langle 0_A\, ,0_B|H_{int}|I\rangle\langle I|H_{int}|0_A\, ,0_B\rangle}{E_I-(E_0^A+E_0^B)} \, .
\end{equation}
The prime indicates that we must sum over all possible states $|I\rangle\neq|0_A\, ,0_B\rangle$. $E_0^A$ and
$E_0^B$ are the energies of the ground states of the atoms $A$ and $B$, respectively.
Denoting the possible states of atom $A$ by $|r\rangle$ and the possible states of atom $B$ by $|s\rangle$,
we may write the above formula as
\begin{equation}
	\label{pertsegord}
	E_{NR}^{(2)}(\mathbf{r}_A,\mathbf{r}_B)=-\sum\limits_{r,s}\,\!'\,\, \frac{\langle 0_A\, ,0_B|H_{int}|r,s\rangle\langle r,s|H_{int}|0_A\, ,0_B\rangle}{E^{A}_{0r}+E^{B}_{0s}} \, ,
\end{equation}
where $E^{A}_{0r}=E^A_{r}-E^A_{0}$ and $E^{B}_{0s}=E^B_{s}-E^B_{0}$.
Equation (\ref{hintdois}) shows that in principle we have 16 terms to deal with in eq. (\ref{pertsegord}) but
fortunately as we 
shall see, most of them either vanish or are irrelevant for our purposes.
For the sake of clarity we analyse them separately. 
To begin with, the term
\begin{eqnarray}
	E_A^{(2)}(\mathbf{r}_A)\!\!\!&=&\!\!\!-\sum\limits_{r,s}\,\!'\,\, \frac{\langle 0_A\, ,0_B|H_A|r,s\rangle\langle r,s|H_A|0_A\, ,0_B\rangle}{E^{A}_{0r}+E^{B}_{0s}} \cr\cr
 \label{eui2}&=& -\sum\limits_{r}\,\!'\,\, \frac{\langle 0_A|H_A|r\rangle\langle n|H_A|0_A\rangle}{E^{A}_{0r}} \, ,
\end{eqnarray}
since only intermediate states with $|s\rangle =|0_B\rangle$ survive in the summation. This term depend only on atom $A$
and stands for 
the second-order contribution to the interaction between the atom $A$ and the surface. An analogous term 
holds for atom $B$. Since these two terms doesn't contribute to non-additivity we shall henceforth neglect them.

%
%

%
%
%
%
%
%
%
All the other 10 terms involving $H_i$
vanish. Indeed, the crossed term involving $H_A$ and $H_B$,
\begin{equation}
	-\sum\limits_{r,s}\,\!'\,\, \frac{\langle 0\, ,0|H_A|r,s\rangle\langle r,s|H_B|0\, ,0\rangle}{E^{A}_{0r}+E^{B}_{0s}}=0 \label{hahb}
\end{equation}
because the term $\langle 0\, ,0|H_A|r,s\rangle$ isn't zero only for intermediate
states with $s=0$, while $\langle r,s|H_B|0,0\rangle$ vanishes in such cases. From eq. (\ref{vabs}) we
see that the same argument 
applies to the crossed term  
\begin{eqnarray}
&-&\sum\limits_{r,s}\,\!'\,\, \frac{\langle 0\, ,0|H_A|r,s\rangle\langle r,s|H_{ABS}|0\, ,0\rangle}{E^{A}_{0r}+E^{B}_{0s}}=\cr\cr
 &-&\frac{1}{\varepsilon_0}\sum\limits_{r,s}\,\!'\,\,\frac{\langle 0\, ,0|H_A|r,s\rangle\langle r,s|d^A_id^B_j|0\, ,0\rangle}{E^{A}_{0r}+E^{B}_{0s}}\times\cr\cr
 &\times&\nabla_i\nabla_j'G_H(\mathbf{r},\mathbf{r}')\bigg|_{\mathbf{r}=\mathbf{r}_A,\mathbf{r}'=\mathbf{r}_B}\, . \label{hahabs}
\end{eqnarray}
Summing up,  equation (\ref{pertsegord})
becomes
\begin{eqnarray}
	\label{unr2}
E_{NR}^{(2)}(\mathbf{r}_A,\mathbf{r}_B)&=&E_A^{(2)}(\mathbf{r}_A)+E_{B}^{(2)}(\mathbf{r}_B)+\cr\cr
&&\!\!\!\!\!\!\!\!\!\!+ E_{Lon}(\mathbf{r}_A,\mathbf{r}_B)+E_{NA}(\mathbf{r}_A,\mathbf{r}_B) \, ,
\end{eqnarray}
where
\begin{eqnarray}
	E_{Lon}&=&-\sum\limits_{r,s}\,\!'\,\,\frac{\langle 0\, ,0|H_{AB}|r,s\rangle\langle r,s|H_{AB}|0\, ,0\rangle}{E^{A}_{0r}+E^{B}_{0s}} \nonumber \\ \label{london}
\end{eqnarray}
and
\begin{eqnarray}
E_{NA}&=&-\sum\limits_{r,s}\,\!'\,\,\frac{\langle 0\, ,0|H_{AB}|r,s\rangle\langle r,s|H_{ABS}|0\, ,0\rangle}{E^{A}_{0r}+E^{B}_{0s}}+\cr\cr
&-&\sum\limits_{r,s}\,\!'\,\,\frac{\langle 0\, ,0|H_{ABS}|r,s\rangle\langle r,s|H_{AB}|0\, ,0\rangle}{E^{A}_{0r}+E^{B}_{0s}} +\cr\cr
&-&\sum\limits_{r,s}\,\!'\,\,\frac{\langle 0\, ,0|H_{ABS}|r,s\rangle\langle r,s|H_{ABS}|0\, ,0\rangle}{E^{A}_{0r}+E^{B}_{0s}} \, . \label{elondoneabs}
\end{eqnarray}
Hence, we are left with just four terms to analyze. The one contained in $E_{Lon}$ is independent of
the surface and the only one that survives in the absence of the surface ($G_H=0$), so it
accounts for the interaction between the atoms in the vacuum. The other three terms, expressed
in $E_{NA}$, are the
key players in this paper and contain the non-additivity effects.
Before delving deeper in this contribution, though, 
let us focus on $E_{Lon}$ in order to re-write it in a more familiar
way. From equation (\ref{vabs}) we see that 
\begin{eqnarray}
	H_{AB}&=&\frac{1}{4\pi\varepsilon_0}(d^A_id^B_j)\nabla_j'\nabla_i\frac{1}{|\mathbf{r}-\mathbf{r}'|}\Bigg|_{\mathbf{r}=\mathbf{r}_A,\mathbf{r}'=\mathbf{r}_B} \cr\cr
&=& \frac{d^A_id^B_j}{4\pi\varepsilon_0 R_{AB}^3} \left(\delta_{ij}-3(\hat{R}_{AB})_i(\hat{R}_{AB})_j\right) \, , \label{londonR}
\end{eqnarray}
where $\mathbf{R}_{AB}:=\mathbf{r}_A-\mathbf{r}_B$. Therefore, substituting last equation into (\ref{london}) we get
\begin{eqnarray}
	E_{Lon}&=&-\sum\limits_{r,s}\,\!'\,\,\frac{\langle 0\, ,0|d^A_id^B_j|r,s\rangle\langle r,s|d^A_md^A_n|0\, ,0\rangle}{(4\pi\varepsilon_0)^2 R_{AB}^6(E^{A}_{0r}+E^{B}_{0s})} \times \cr\cr
&& \!\!\!\!\!\!\!\!\!\!\!\!\!\!\!\!\!\!\!\!\! \left(\delta_{ij}-3(\hat{R}_{AB})_i(\hat{R}_{AB})_j\right)\left(\delta_{mn}-3(\hat{R}_{AB})_m(\hat{R}_{AB})_n\right) \, . \nonumber \\ \label{londonincompleto}
\end{eqnarray}
As usual, we adopt the notation $\langle 0 |d^A_i|r\rangle = d^{or}_i$ and, for the atom $B$,  
$\langle 0 |d^B_j|s\rangle = d^{os}_j$. Therefore we write
\begin{equation}
\langle 0\, ,0|d^A_id^B_j|r,s\rangle\langle r,s|d^A_md^A_n|0\, ,0\rangle=d^{0r}_id^{r0}_md^{0s}_jd^{s0}_{n} \, .
\end{equation}
As we are dealing with 
freely rotating atoms, the above transition elements must be averaged over all directions. Furthermore,
assuming isotropy of the atoms, we have
\begin{equation}
	\overline{d^{0r}_id^{r0}_m}=\delta_{im}\frac{|\mathbf{d}^{or}_A|^2}{3} \, , \label{sphersy}
\end{equation}
where the symbol $\overline{\cdots}$ denotes the average over all directions. 
An analogous equation holds for atom $B$. So, collecting the last two equations and substituting them into (\ref{londonincompleto})
we obtain London's result in its most common form\cite{Thiru}, 
\begin{eqnarray}
	E_{Lon}&=&-\frac{1}{24\pi^2\varepsilon_0^2 R_{AB}^6}\sum\limits_{r,s}\,\!'\,\,\frac{|\mathbf{d}^{or}_A|^2|\mathbf{d}^{0s}_B|^2}{(E^{A}_{0r}+E^{B}_{0s})} \, . \label{londonfinal}
\end{eqnarray}
Having re-obtained this important expression as a particular case of our general expressions, we turn to
the non additive terms in next section.


\section{The non-additivity term}

Let us now concentrate our attention in the non additive effects. Equation (\ref{elondoneabs}) is the 
only one that depends
simultaneously on both atoms and the surface - it reflects and contains the non-additivity effects that are inherent to
the van der Waals dispersive interaction. 
The mathematical treatment of this term is completely analogous to that given to $E_{Lon}$
in the last section. In such a way, the first two terms on the right-hand-side of eq.(\ref{elondoneabs}) are 
equal and given by
\begin{eqnarray}
	&&-\sum\limits_{r,s}\,\!'\,\,\frac{\langle 0\, ,0|H_{AB}|r,s\rangle\langle r,s|H_{ABS}|0\, ,0\rangle}{E^{A}_{0r}+E^{B}_{0s}}= \cr\cr
&=& -\sum\limits_{r,s}\,\!'\,\,\frac{\langle 0\, ,0|H_{ABS}|r,s\rangle\langle r,s|H_{AB}|0\, ,0\rangle}{E^{A}_{0r}+E^{B}_{0s}} =\cr\cr
&=& -\frac{1}{36\pi\varepsilon_0^2 R_{AB}^3}\sum\limits_{r,s}\,\!'\,\,\frac{|\mathbf{d}^{or}_A|^2|\mathbf{d}^{0s}_B|^2}{(E^{A}_{0r}+E^{B}_{0s})}\bigg\{\mathcal{G}^{H}_{ii}(\mathbf{r}_A,\mathbf{r}_B)+\cr\cr
&-& 3(\hat{R}_{AB})_i(\hat{R}_{AB})_j\mathcal{G}^{H}_{ij}(\mathbf{r}_A,\mathbf{r}_B)\bigg\} \, ,
 \label{termo1}
\end{eqnarray}  
while the last term on the right-hand-side of eq.(\ref{elondoneabs}) is
\begin{eqnarray}
	&&-\sum\limits_{r,s}\,\!'\,\,\frac{\langle 0\, ,0|H_{ABS}|r,s\rangle\langle r,s|H_{ABS}|0\, ,0\rangle}{E^{A}_{0r}+E^{B}_{0s}}= \cr\cr
&=&-\frac{1}{9\varepsilon_0^2}\sum\limits_{r,s,i,j}\,\!'\,\,\frac{|\mathbf{d}^{or}_A|^2|\mathbf{d}^{0s}_B|^2}{(E^{A}_{0r}+E^{B}_{0s})} [\mathcal{G}^{H}_{ij}(\mathbf{r}_A,\mathbf{r}_B)]^2\, ,\nonumber\\ \label{termo2}
\end{eqnarray}
where we defined
\begin{equation}
		\mathcal{G}^{H}_{ij}(\mathbf{r}_1,\mathbf{r}_2)=\nabla_i\nabla_j'G_H(\mathbf{r},\mathbf{r}')\bigg|_{\mathbf{r}=\mathbf{r}_1,\mathbf{r}'=\mathbf{r}_2} \, . \label{gij}
\end{equation}
Substituting eqs. (\ref{termo1}-\ref{termo2}) into 
eq.(\ref{elondoneabs}) we obtain

\begin{equation}
	E_{NA}=E_{NA}^{(1)}+E_{NA}^{(2)} \, ,
\end{equation}

with
\begin{eqnarray}
	E_{NA}^{(1)}&=& -\frac{\Lambda_{AB}}{18\pi\varepsilon_0^2 R^3} \Bigg[\mbox{Tr}\,\mathcal{G}^{H}(\mathbf{r}_A,\mathbf{ r}_B)+  \cr\cr
&-&3(\hat{R}_{AB})_i(\hat{R}_{AB})_j\mathcal{G}^{H}_{ij}(\mathbf{r}_A,\mathbf{r}_B)\Bigg] \label{ena1} \\
   E_{NA}^{(2)}&=&  -\frac{\Lambda_{AB}}{9\varepsilon_0^2}\sum\limits_{i,j}[\mathcal{G}^{H}_{ij}(\mathbf{r}_A,\mathbf{r}_B)]^2\, , \label{ena}
\end{eqnarray}
where 
\begin{eqnarray}
	\Lambda_{AB}&=&\sum\limits_{r,s}\,\!'\,\,\frac{|\mathbf{d}^{or}_A|^2|\mathbf{d}^{0s}_B|^2}{(E^{A}_{0r}+E^{B}_{0s})} \, . \label{Lambda}
\end{eqnarray}
Eqs.(\ref{ena1}) and (\ref{ena}) constitute the main result of this paper.
We now proceed to evaluate the influence exerted by the surface on the interatomic interaction by analysing the
ratio 
\begin{eqnarray}
&&	\frac{E_{NA}}{E_{Lon}}=\frac{4\pi R_{AB}^3}{3} \Bigg [\mbox{Tr}\,\mathcal{G}^{H}(\mathbf{r}_A,\mathbf{r}_B)-3(\hat{R}_{AB})_i(\hat{R}_{AB})_j\cr\cr
&\times&\mathcal{G}^{H}_{ij}(\mathbf{r}_A,\mathbf{r}_B)\Bigg]+ \frac{8\pi^2 R_{AB}^6}{3}\sum\limits_{i,j}\Bigg [\mathcal{G}^{H}_{ij}(\mathbf{r}_A,\mathbf{r}_B)\Bigg]^2 \, . \label{ratio}
\end{eqnarray}
The influence of the surface on the atomic interaction, in this order of perturbation theory, is a purely geometrical effect,
since it doesn't depend on the internal structure of the atoms. The formulas developed in this section show that in order to calculate the non-addivity
effects all we have to do is to find $G_H$, which can be done by mapping our problem
into an electrostatic one, according to the eq.(\ref{ghimages}). In the next section 
we will illustrate this method with treating some examples. 
 

\section{Applications\label{applications}}

\subsection{Two atoms and a conducting infinite plane}

\begin{figure}[!ht]
\centering
\includegraphics[scale=0.60]{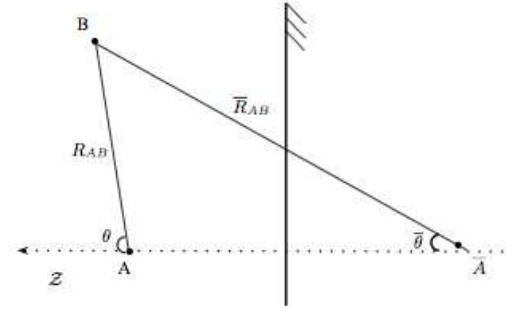}
\caption{(color online) Two atoms, A and B, in the presence of an infinite conducting plane. $\overline{A}$ is the image of A and the image of $B$ is not represented. $R_{AB}$
	is the distance between the atoms and $\overline{R}_{AB}$ is the distance between $B$ and the image of $A$,
	denoted by $\overline{A}$.}
\label{doisatomosplano}
\end{figure}

As a first example let us consider two atoms in front an the influence of an infinite conducting plane as illustrated
in figure \ref{doisatomosplano}.

We choose the axis as to have the conducting plane at $z=0$ and both atoms belonging to the $\mathcal{XZ}$ plane.
By eq. (\ref{ghimages}) we get
\begin{equation}
	G_H(\mathbf{r},\mathbf{r}')=-\frac{1}{4\pi\sqrt{(x-x')^2+(y-y')^2+(z+z')^2}} \, .
\end{equation}
Employing eq. (\ref{gij}), the only non vanishing terms are
\begin{eqnarray}
	\mathcal{G}_{xx}&=&\frac{3\sin^2\overline{\theta}-1}{4\pi\overline{R}_{AB}^3} \\
	\mathcal{G}_{yy}&=&-\frac{1}{4\pi\overline{R}_{AB}^3} \\
	\mathcal{G}_{zz}&=&\frac{1-3\cos^2\overline{\theta}}{4\pi\overline{R}_{AB}^3} \\
	\mathcal{G}_{xz}&=&\frac{3\sin\overline{\theta}\cos\overline{\theta}}{4\pi\overline{R}_{AB}^3}=-\mathcal{G}_{zx} \, ,
\end{eqnarray}
where we used $R_{AB}\sin\theta=\overline{R}_{AB}\sin\overline{\theta}$. Substituting these expressions into
eq.(\ref{ena}) we arrive at
\begin{equation}
	E_{NA}^{(2)}=-\frac{\Lambda_{AB}}{24\pi^2\varepsilon_0^2 \overline{R}_{AB}^6}=E_{Lon}(\overline{R}) \, ,
	\label{londonimage}
\end{equation}
where we used eq. (\ref{londonfinal}). This term stands for the London interactions between an atom
and the image of the other. 
To evaluate the other term, note that $\hat{R}_x=\sin\epsilon$ and $\hat{R}_z=\cos\epsilon$,
while the $y$-component vanishes. Hence eq. (\ref{ena1}) yields
\begin{eqnarray}
	E_{NA}^{(1)}=-\frac{\Lambda_{AB}}{72\pi^2\varepsilon^2 R^3\overline{R}^3}(2-3\sin^2\theta-3\sin^2\overline{\theta}) \, .
\end{eqnarray}
We have now the complete expression to the non-additivity terms up to second order in perturbation theory
 for the two atoms-conducting plane case. The sum $E_{NA}^{(1)}+E_{NA}^{(2)}$ coincides precisely to
the result obtained by Power and Thirunamachandran\cite{power}.

\subsection{Two atoms inside a plane capacitor}

This constitutes the main example of this paper. As we show, the non-additivity effects can are noticeable even at
large interatomic separations as it strongly suppress the interatomic interaction.

%

%
\begin{figure}[!ht]
\centering
\includegraphics[scale=0.60]{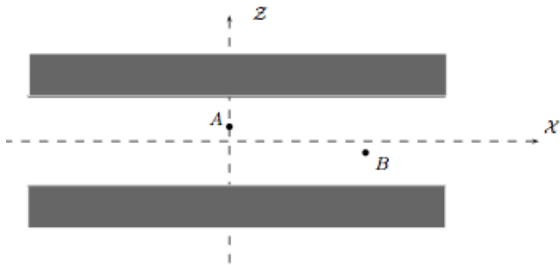}
\caption{(color online) Two atoms, A and B inside a perfectly conductor plate conductor. We choose the the plane $z=0$ midway between
	the plates. Without loosing generality, atom $A$ is put at $(0,0,z_A)$ and $B$ at $(x_B,0,z_B)$.}
\label{doisatomosdoisplanos}
\end{figure}

It is convenient to choose our coordinates in such a way as to have
the plane $z=0$ equidistant to the conducting plates, which are separated by
a distance $D$. We orient the axis to have both atoms in the $\mathcal{X}\mathcal{Z}$ plane as illustrated
in figure \ref{doisatomosdoisplanos}.
The $G_H$ is obtained through the solution of the electrostatic problem of one charge in the presence
of a plane capacitor, which can be done again employing the image method. In this case, however, we must
deal with an infinite series of images. The potential generated by this series, however, is very slowly convergent
and it is more convenient to write it in another form. We follow Ref.\cite{pumplin} to find that for this geometry
the Green function may be written as
\begin{eqnarray}
G(\mathbf{r},\mathbf{r}')\!=\!\frac{1}{\pi D} \sum_{n=1}^{\infty} \cos\frac{n\pi z}{D}\cos\frac{n\pi z'}{D}K_0\!\left(\frac{n\pi|\boldsymbol{\rho}-\boldsymbol{\rho'}|}{D}\right)  \label{g2p}
\end{eqnarray}
%
%
%
where $\boldsymbol{\rho}=x\mathbf{\hat{x}}+y\mathbf{\hat{y}}$ and $\boldsymbol{\rho'}=x'\mathbf{\hat{x}}+y'\mathbf{\hat{y}}$
and $K_0$ is a modified Bessel function of the second kind\cite{especiais}. Note that in the last equation figures the complete
Green function given in eq.(\ref{G}), instead of $G_H$. In this example it pays off to work directly with $G$, and
in the end we turn to $G_H$ in order to isolate the non-additivity contribution. 

The asymptotic expansion of eq.(\ref{g2p}), valid in the region $|\boldsymbol{\rho}-\boldsymbol{\rho'}|\gtrsim D$,
is given by
\begin{eqnarray}
	G(\mathbf{r},\mathbf{r}')=\frac{1}{4\pi}\sqrt{\frac{8}{|\boldsymbol{\rho}-\boldsymbol{\rho'}|D}}\cos\frac{\pi z}{D}\cos\frac{\pi z'}{D}e^{-\frac{\pi|\boldsymbol{\rho}-\boldsymbol{\rho'}|}{D}}  \label{assymptotic}
	\, .
\end{eqnarray}
Working directly with $G$, the four terms depicted in eqs. (\ref{londonincompleto}), (\ref{termo1}) and (\ref{termo2}) can be naturally assembled in just one term
\begin{eqnarray}
	E_{Lon}+E_{NA}=-\frac{\Lambda_{AB}}{9\varepsilon_0^2}\sum\limits_{i,j}[\mathcal{G}_{ij}(\mathbf{r}_A,\mathbf{r}_B)]^2 \, ,
\end{eqnarray}
where 
\begin{eqnarray}
	\mathcal{G}_{ij}(\mathbf{r}_A,\mathbf{r}_B)&=&\frac{1}{4\pi}\nabla_i\nabla_j'\Bigg\{\cos\frac{\pi z}{D}\cos\frac{\pi z'}{D}e^{-\frac{\pi|\boldsymbol{\rho}-\boldsymbol{\rho'}|}{D}} \times \cr\cr
&& \sqrt{\frac{8}{|\boldsymbol{\rho}-\boldsymbol{\rho'}|D}}\Bigg\}\bigg|_{\mathbf{r}=\mathbf{r}_A,\mathbf{r}'=\mathbf{r}_B} \, .
\end{eqnarray}
Hence, when the atoms are separated by a distance of the order of $D$ or larger, the non-addivity effects 
shield the atoms exponentially. This is a remarkable result and it is particularly interesting when both atoms are equidistant from
the plates. In this case, we see immediately by symmetry that the force exerted by the surface in each
atom separately vanishes. Nevertheless, the plates leave their footprint on the interatomic force by suppressing
exponentially an interaction that would fall with $1/R_{AB}^6$ in vacuum. The results
we display here for two atoms remain valid up to second order of perturbation when several atoms are present.
This is so because in this order no term in the interaction energy of the system can couple more than two atoms. This can
be seen by employing a reasoning analogous to that we used to
show the terms in eqs.(\ref{hahb}) and (\ref{hahabs}) vanish.
Hence, if a gas is rarefied enough so that
its atoms are  on average separated by a distance $D$ or more, 
then up to second order of perturbation the capacitor shields 
the interatomic interaction producing an ideal gas behavior. 
Even when this condition is not strictly satisfied our results 
show that the atoms interact only with the atoms closer than $D$, leading us to the expectation 
that a gas put between parallel 
conducting plates behaves more ideally than otherwise. Since the interatomic 
interaction is the main feature responsible
for the gas-liquid phase transition, we conjecture that putting a gas between conducting plates could lower the 
liquefaction temperature. 

\begin{figure}[!ht]
\centering
\includegraphics[scale=0.35]{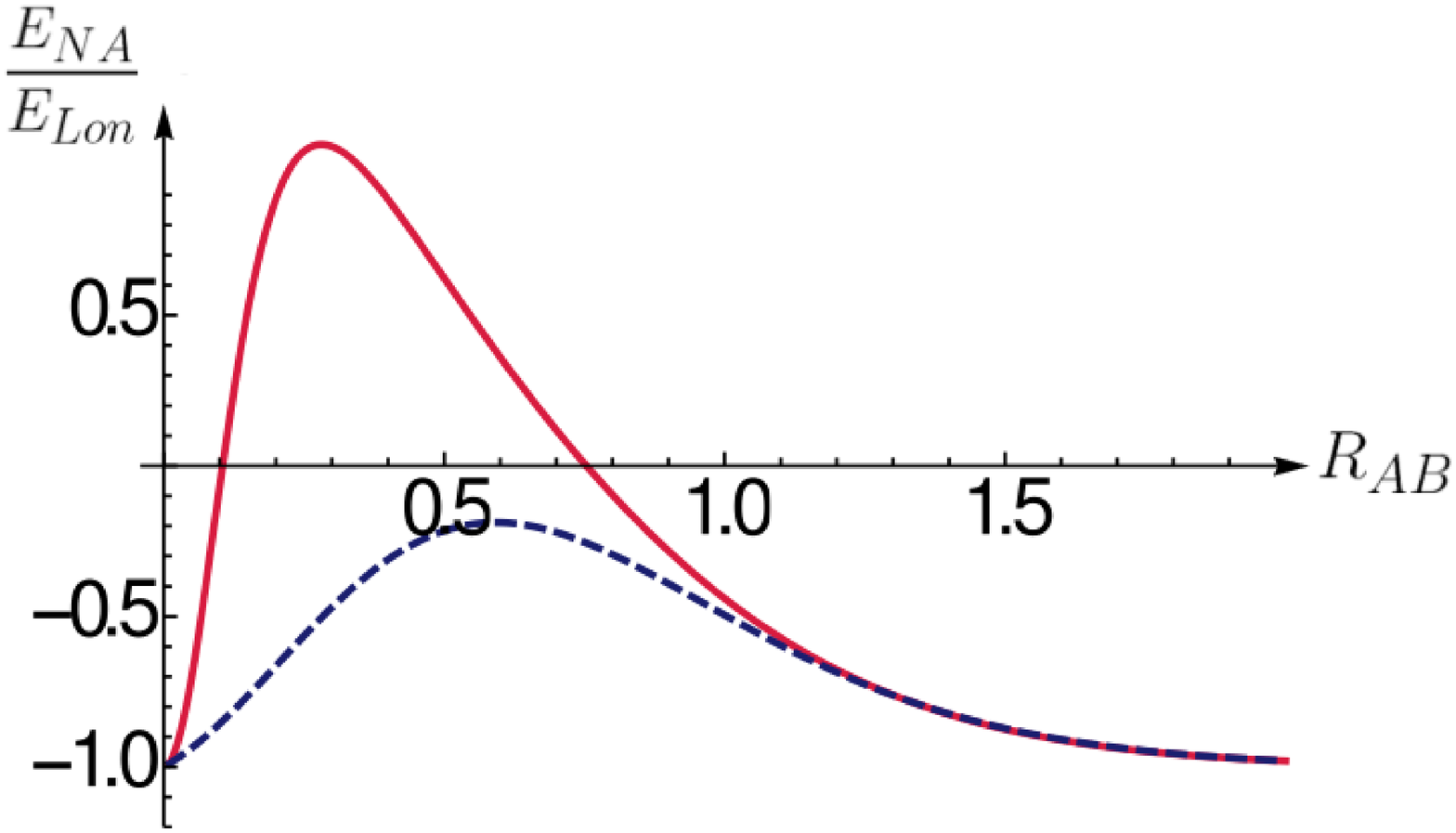}
\caption{(color online) Non-additive part of the interaction energy (normalized by the London ineraction energy) as a function of the distance between atom $A$ and atom $B$. The atoms are both equidistant to the plates. $R_{AB}$ measured in units of $D$. The dashed blue curve employs the asymptotic expression given in eq.(\ref{assymptotic}) and the red
	solid curve employs the complete expression (\ref{g2p}).}
\label{Fig6}
\end{figure}

\begin{figure}[!ht]
\centering
\includegraphics[scale=0.35]{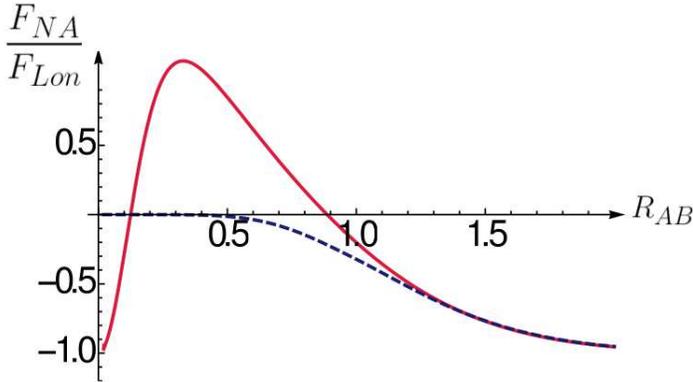}
\caption{(color online) Non-additive part of the force (normalized by the force between the atoms in vacuum) as a function of the distance between atom $A$ and atom $B$. The atoms are both equidistant to the plates. $R_{AB}$ measured in units of $D$. The dashed blue curve employs the asymptotic expression given in eq.(\ref{assymptotic}) and the red solid curve employs the complete expression (\ref{g2p}).}
\label{Fig7}
\end{figure}

At last, we isolate the non-additivity effects through equations (\ref{ena1}) and (\ref{ena}). In order to obtain
$G_{H}$ we must subtract $1/|\mathbf{r}-\mathbf{r}'|$ from eq.(\ref{g2p}). In figure \ref{Fig6} 
we plot the non-additivity effects normalized
by the London energy for both the complete expression and the asymptotic expansion (obtained using eq.(\ref{assymptotic}) for $G_H$), in the situation where both atoms
are $z=0$ (equidistant from the conducting planes).
As expected from our previous discussion, for large distances the ratio goes to $-1$, showing that the non-additivity
cancels out the London interaction between the atoms. Finally, in figure \ref{Fig7} we do the same thing 
for the $\rho$-component of the force on atom $A$.  

\subsection{Two atoms and a conducting grounded sphere}

In this section we analyse the non-retarded interaction between two atoms and a grounded
perfectly conducting sphere. 
Since the application of the image method to the sphere is well-known, we shall use it to obtain $G_H$.
Placing the origin of our coordinates at
the center of the sphere of radius $a$ and setting $\mathbf{r'}$ as the position of the physical charge,
 we have to put an image charge $q_{i} = -\frac{a}{r^{\,\prime}}q$ at position 
 $\mathbf{r}^{\,\prime}_{i} = \frac{a^2}{r'^2}\mathbf{r}'$\cite{griffiths}. Therefore, from eq. (\ref{ghimages}) we have
 \begin{equation}
G_H(\mathbf{r},\mathbf{r}')=-\frac{a}{4\pi r^{\,\prime}|\mathbf{r}-\mathbf{r}^{\,\prime}_{i}|} =-\frac{a}{4\pi\sqrt{r^2r'^2-2\mathbf{r}\cdot\mathbf{r}'a^2+a^4}} \, . \label{ghsphere}
\end{equation}
Following the scheme outlined in the last sections, we now use this function to analyse the quantum dispersive 
interaction between the atoms and the sphere. Calculating $\mathcal{G}_{ij}$ from eq. ($\ref{gij}$),
we obtain after some algebra
\begin{eqnarray}
	\mathcal{G}^{H}_{ij} &=&-\frac{3a(x^A_ir_B^2-x^B_ia^2)(x^B_jr_A^2-x^A_ja^2)}{4\pi[r_A^2r_B^2-2\mathbf{r}_A\cdot\mathbf{r}_Ba^2+a^4]^{5/2}}+ \cr\cr
&+&\frac{a(2x^A_ix^B_j-\delta_{ij}a^2)}{4\pi[r_A^2r_B^2-2\mathbf{r}_A\cdot\mathbf{r}_Ba^2+a^4]^{3/2}} \, , \label{gijesfera}
\end{eqnarray}
where $x^{A}_i$ stands for the $i$-th cartesian coordinate of $\mathbf{r}_A$, $r_A=|\mathbf{r}_A|$, with analogous
notation for the coordinates of $B$.
It is convenient to orient the axis in order to have the atom $A$ at $(0,0,r_A)$ and the atom $B$ at $(0,r_B\sin\theta,r_B\cos\theta)$. Employing equations (\ref{ena}) we get the complete expression for the non-additivity terms for
any configuration of the atoms, but for the sake of clarity we write explicitly only two particular cases.
When the two atoms are aligned with the center of the sphere, we see from (\ref{gijesfera}) that
$\mathcal{G}$ is diagonal. In that case, we have $\theta=0$ ($\theta=\pi$) when the atoms are on the 
same side (opposite side) of the sphere, thus the non-additivity terms are given by
\begin{eqnarray}
E_{NA}^{(1)}&=& \pm\frac{\Lambda_{AB}}{36\pi^2\varepsilon_0^2 R_{AB}^3}\frac{ar_Ar_B}{(r_Ar_B\pm a^2)^3} \\
E_{NA}^{(2)}&=& -\frac{\Lambda_{AB}}{144\pi^2\varepsilon_0^2}\frac{3a^6\mp2a^4r_Ar_B+a^2r_A^2r_B^2}{(r_Ar_B\pm a^2)^6}\, , \label{enaesfera}
\end{eqnarray}
where the upper (lower) sign refers to the $\theta=\pi$ ($\theta=0$) case. Note that both terms
are bigger for $\theta=0$. Since, by symmetry, we expect the same behaviour when we change $\theta$ by
$2\pi-\theta$, we conclude that $\theta=\pi$ will be a minimum for the non-additivity interaction energy. 

\begin{figure}[!ht]
\centering
\includegraphics[scale=0.60]{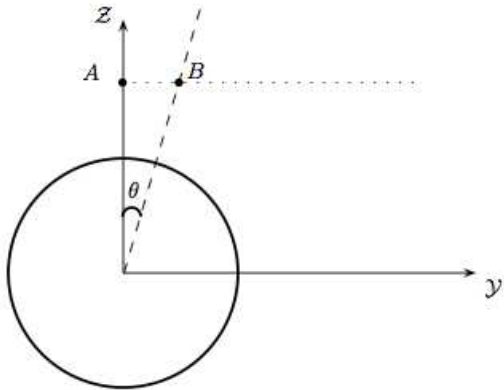}
\caption{Two atoms near a conducting sphere. $\mathbf{R}_{AB}$ is parallel to the $\mathcal{X}$-axis. }
\label{exesfera}
\end{figure}

In this case, in contrast with the capacitor example, non-additivity effects are practically unnoticeable since by far
the dominant contribution comes by far from the attraction of each atom with the sphere, which contributes in first
order of perturbation theory. However, since symmetry considerations ensure that the force exerted by the sphere
on each atom separately is radial, non-additivity may be relevant in the components of the force
transverse to that direction. Let us then consider the situation
where the atom $A$ is at the $\mathcal{Z}$-axis and $\mathbf{R}_{AB}$ is parallel to the $\mathcal{Y}$
direction, as illustrated in figure \ref{exesfera}.
%
%
%

%

\begin{figure}[!h]
\centering
\includegraphics[scale=0.45]{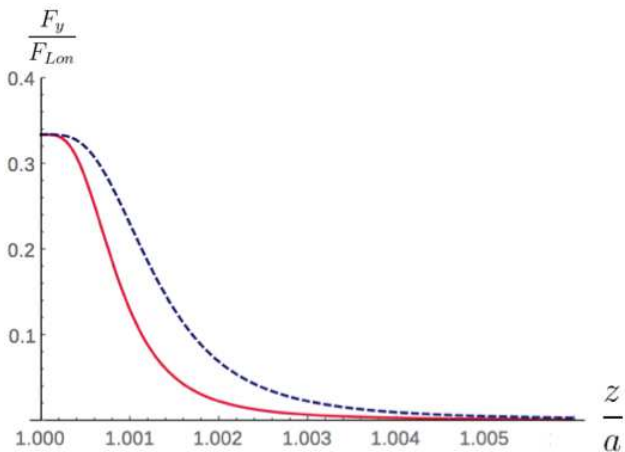}
\caption{(color on-line) Non-additive part of the x-component of the force exerted on atom $B$ (normalized by the London force) as a function of the distance between atom $B$ and the center of the sphere. $\mathbf{R}_{AB}$ remains always perpendicular
	to $\mathbf{r}_B$. The red solid curve is for two atoms separated by $d_A = 0.002 a$ and the dashed blue curve is for $d_A = 0.003 a$.}
\label{Fig5}
\end{figure}

In this setup, only the London interaction and non-additivity terms 
contribute to the $y$-component of the force on atom $B$. In figure \ref{Fig5} we 
plot this component of the force as a function of the distance from atom $B$ to the center of the sphere,
keeping the vector $\mathbf{R}_{AB}$ fixed. 
We see that the non-additivity parcel may indeed be comparable to London interaction for close distances. For a sphere of $1 \mu$m of radius, and the atom $B$ at a distance of $1$ nm from the
surface of the sphere, and separated by a distance $2$ nm from atom $A$, the non-additivity force is $30\%$ from
the London force between the atoms.

To end this example we make some last remarks. First, remember that in our first example
of two atoms in the presence of a plane, $E_{NA}^{2}$, given in eq.(\ref{londonimage}), could readily be
 identified as the London interaction between each atom and the image of the other. 
 In this example, $E_{NA}^{(2)}$ does not have such a simple interpretation. This is related to the fact
 that to solve the electrostatic problem of a dipole in the presence of a grounded conducting sphere we must have not
 only an image dipole but two point charges as well\cite{fila}. As a particular case of our results
 we may obtain, by substituting eq.(\ref{ghsphere}) into eq.(\ref{ui}) the dispersive interaction 
 between one atom and a conducting sphere. In doing so we arrive at the same result obtained in Ref.\cite{taddei}.

\subsection{Two atoms and a conducting isolated neutral sphere}

We need only minor modifications to tackle the case where the two atoms are in the presence 
of an isolated, instead of a grounded sphere. In the electrostatic case, the isolated sphere
interacts weaklier with a charge, since in the grounded case the sphere is supplied by the earth with
additional charges. Let us see what happens to the non-additivity effects in this quantum problem.

\begin{figure}[!h]
\centering
\includegraphics[scale=0.31]{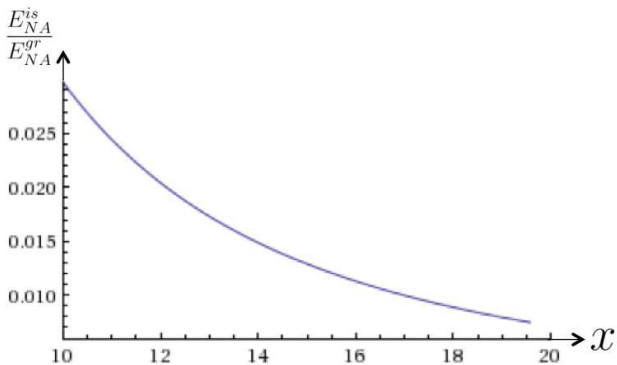}
\caption{Ratio between the non-additive the isolated sphere's and the grounded sphere's non additive contributions 
	for the interaction energy. The horizontal axis stands for the distance of atom $A$ to the center of the sphere (in units of the radius of the sphere). The atom $B$
is farther from the sphere, separated by a distance $R_{AB}=0.002a$ from atom $A$ and co-linear with it and 
the center of the sphere.}
\label{isogro}
\end{figure}

$G_H$ is obtained from the one in the last case by adding only one term\cite{nosajp}, namely,
\begin{equation}
	G_H(\mathbf{r},\mathbf{r}')=-\frac{a}{4\pi\sqrt{r^2r'^2-2\mathbf{r}\cdot\mathbf{r}'a^2+a^4}}+\frac{a}{4\pi rr'} \, .\label{ghi}
\end{equation}
Substituting eq.(\ref{ghi}) into eq.(\ref{gij}) and using our previous result, eq. (\ref{gijesfera}), we obtain
\begin{eqnarray}
	\mathcal{G}^{H}_{ij} &=&-\frac{3a(x^A_ir_B^2-x^B_ia^2)(x^B_jr_A^2-x^A_ja^2)}{4\pi[r_A^2r_B^2-2\mathbf{r}_A\cdot\mathbf{r}_Ba^2+a^4]^{5/2}}+ \cr\cr
+&& \!\!\!\!\!\!\!\frac{a(2x^A_ix^B_j-\delta_{ij}a^2)}{4\pi[r_A^2r_B^2-2\mathbf{r}_A\cdot\mathbf{r}_Ba^2+a^4]^{3/2}} +\frac{ax_i^Ax_j^B}{4\pi r_A^3r_B^3} \label{gijesferais}
\end{eqnarray}
Once more, we may evaluate the non-additivity contribution to the interaction energy of the system in any of its configurations. As before, we write explicitly
only the case when the atoms are co-linear with the center of the sphere.
Using eqs. (\ref{enaesfera}), we obtain
\begin{eqnarray}
E_{NA}^{(1)}&=& \pm\frac{\Lambda_{AB}}{36\pi^2\varepsilon_0^2 R_{AB}^3}\left[\frac{ar_Ar_B}{(r_Ar_B\pm a^2)^3}-\frac{a}{r_A^2r_B^2}\right] \\
E_{NA}^{(2)}&=& -\frac{\Lambda_{AB}}{144\pi^2\varepsilon_0^2}\Bigg[\left(\frac{ar_Ar_B\mp a^3}{(r_Ar_B\pm a^2)^3}-\frac{a}{r_{A}^2r_{B}^2}\right)^2+ \nonumber \\
&& \;\;\;\;\;\;\;\;\;\; \; \; + \frac{2a^6}{(r_Ar_B\pm a^2)^6}\Bigg] \, .
\end{eqnarray}
It can be seen in figure \ref{isogro} that the non-additivity effects are greatly diminished in comparison with
the previous case, where the sphere was grounded. 

Let us examine an interesting particular case. Letting $a\longrightarrow 0$, we see
that the first contribution to $E_{NA}^{(1)}$($E_{NA}^{(2)}$) is of third (fourth) order in $a$. 
 Therefore, keeping only the lowest order contribution, we may write for 
the non-additivity term of two atoms plus a very small sphere the expression
\begin{equation}
	E_{NA}=-\frac{K\Lambda_{AB}}{R_{AB}^3r_A^3r_B^3}a^3+\mathcal{O}(a^4) \, , \label{axil}
\end{equation}
where $K$ is a positive constant. In this limit we may identify the shrunk sphere as an atom 
with polarizability $\alpha(C)$ proportional to $a^3$. Also, $r_A$ and $r_B$ become
the respective distance from the atoms $A$ and $B$ to the atom $C$. 
Therefore, we see that our result is compatible to the well known 
Axilrod-Teller potential for three atoms, with the expected negative sign\cite{Axilrod}.
Note that such compatibility follows from the absence of the powers $a^0$, $a$ and $a^2$ in the above expansion, which
didn't happen for the grounded sphere in the last section. As it should, since this limit on the 
grounded sphere can by no means be thought of as an atom.
We may yet present eq.(\ref{axil}) in a more familiar form. For two-level atoms we 
have\cite{Thiru} $\Lambda_{AB}$ proportional
to $\alpha(A)\alpha(B)$, leaving eq. (\ref{axil}) in the form
\begin{equation}
		E_{NA}=-\frac{K \alpha(A)\alpha(B)\alpha(C)}{R^3r_A^3r_B^3}+\mathcal{O}(a^4) \, . \label{axil}
\end{equation}

\section{Final remarks}

We have dealt with two atoms in the presence of a conducting surface of an arbitrary shape. 
In systems composed by three bodies
it is well-known that non-additivity effects must be taken into account. In the case of three atoms
the effects are rather small since they are of third order in perturbation theory while the additivity
terms are of second order. When a conducting surface is present, however, the non-additivity term is
also of second order and may be relevant to the interaction between the atoms. We have
obtained explicit expressions that allow us to evaluate analytically such an influence from
the solution of a considerably simpler electrostatic problem of only one charge with the conducting
surface, which may be treated by image method whenever possible. We checked the self consistency of our
results re-obtaining London formula. Besides, for the case of two atoms and a conducting plane we also
re-obtained the result displayed in literature. Then we discussed our most important example of two atoms
inside a plate capacitor, where the non-additivity can not be neglected. We showed that the non-additivity
shields one atom from the other, making the interaction between them to fall exponentially with the distance. 
This effect is present also for $N$ atoms between two infinite planes up to second order in perturbation theory.
In such a way we concluded that a gas is closer to an ideal one between conducting planes, 
leading us to conjecture that the gas-liquid transition takes place at lower temperatures 
inside a plate capacitor than inside non-conducting plates. As a last example we treated two atoms in a presence
of a conducting sphere both grounded and isolated. We demonstrated that when the sphere isolated the non-additivity
is much smaller than in the grounded case. The isolated case, however, has a nice particular limit, namely, the three atoms configuration, obtained when we let the radius of the sphere go to zero. 
We expect that the general and simple nature of the results enlisted in this paper allows for a broader understanding of non-additivity, in situations 
where the distances involved are small enough to a non-retarded treatment be appropriate.

%
%
{\bf Acknowledgements}\\
The authors are indebted with P.A. Maia Neto, Kimball A. Milton and Prachi Parashar for valuable discussions. The authors  also  thank to CNPq and FAPERJ (brazilian agencies) for
partial financial support.

%
%

%

\begin{thebibliography}{99}
%
\bibitem {israel} J.N. Israelachvili, Intermolecular and Surface Forces, Academic Press. (2011)

\bibitem{Milonni} P.W. Milonni, The quantum vacuum, Academic Press (1994).

\bibitem{langbein} D. Langbein, Theory of van der Waals Attraction, Springer (2013).

\bibitem{margeneau} H. Margeneau and N.R. Kestner, Theory of Intermolecular Forces, Pergamon Press, Oxford, (1971).

\bibitem{mahanty} J. Mahanty and B.W. Ninham, Dispersion Foces, Academic Press Inc, London (1977).

\bibitem {Axilrod} B.M. Axilrod and E. Teller, {\it J. Chem. Phys.} {\bf 11}, 299 (1943).

\bibitem{ajp} C. Farina, F.C. Santos and A.C. Tort, Am.J.Phys. {\bf 67}, 344 (1999).

\bibitem{efimov} V. Efimov, Phys.Lett.B {\bf 33}, 563 (1970).

\bibitem{Kraemer} T. Kraemer et al., Nature, {\bf 440}, 315 (2006).

\bibitem{Pires} R. Pires et al., Phys.Rev.Lett. {\bf 112}, 250404 (2014).

\bibitem{Huang} B. Huang et al., Phys.Rev.Lett. {\bf 112}, 190401 (2014).

\bibitem{braaten} E. Braaten and H.-W. Hammer, Physics Reports, {\bf 428}, 259 (2006).

\bibitem{roy} S. Roy et al., Phys.Rev.Lett., {\bf 111}, 053202 (2013).

\bibitem{pollack} S.E. Pollack, D. Dries, R.G. Hulet, Science, {\bf 326}, 1683 (2009).

\bibitem{kievsky} M. Gattobigio and A. Kievsky, Phys. Rev.A {\bf 90}, 012502 (2014).

\bibitem{lukin} M.D. Lukin et al., Phys. Rev. Lett {\bf 87}, 037901 (2001).

\bibitem{comparat} D. Comparat and P. Pillet, J. Opt. Soc. Am. B {\bf 27}, A208 (2010).

\bibitem{pohl} T. Pohl and P.R. Berman, Phys.Rev.Lett. {\bf 102}, 013004 (2009).

\bibitem{f} E. Shahmoon, I. Mazets and G. Kurizki, PNAS {\bf 111}, 10485 (2014).

\bibitem{mcbla} A.D. McLachlan, Mol.Phys. {\bf 6}, 423 (1963).

\bibitem{mcbla2} A.D. McLachlan, Mol.Phys. {\bf 7}, 381 (1964).

\bibitem{Milton} Kimball A. Milton, Elom Abalo, Prachi Parashar, K.V. Shajesh, Nuovo Cimento C {\bf 36}, 192 (2013).

\bibitem{Milton2} Kimball A. Milton et al. arXiv:1502.06129 (2015).

\bibitem{Eberlein2007} C. Eberlein and R. Zietal,  Phys.Rev. {\bf A 75}, 032516 (2007).

\bibitem{Eberlein2009} C. Eberlein and R. Zietal, Phys. Rev. A {\bf 80}, 012504 (2009).

\bibitem{Eberlein2011} C. Eberlein and R. Zietal, Phys. Rev. A {\bf 83}, 052514 (2011).

\bibitem{Eberlein 2012} C. Eberlein and R. Zietal, Phys. Rev. A {\bf 86}, 052522 (2012).

\bibitem{nosajp} Reinaldo de Melo e Souza, W.J.M. Kort-Kamp, C. Sigaud and C. Farina, Am.J.Phys. {\bf 81}, 366 (2013). 

\bibitem{nos} Reinaldo de Melo e Souza, W.J.M. Kort-Kamp, C. Sigaud and C. Farina, Phys.Rev. A {\bf 84}, 052513 (2011).

\bibitem{f2} E. Shahmoon and G. Kurizki, Phys. Rev. A {\bf 87}, 062105 (2013).

\bibitem{hinds} C.I. Sukenik, M.G. Boshier, D. Cho, V. Sandoghdar and E.A. Hinds, Phys.Rev.Lett, {\bf 70},
560, 1993.

\bibitem{cohen} C. Cohen-Tannoudji, B. Diu and F. Lalo\"e,
 M\'ecanique quantique II. Paris: 
Hermann (1997) Of special interest to this paper is complement $C_{XI}$. 

\bibitem{stakgold} I. Stakgold, Green's Functions and Boundary Value Problem, pg. 200, John Wiley \& Sons (1979).

\bibitem{self} Note that to obtain the interaction energy from eq.(\ref{ui}) one must discard 
the self-interaction present in the integral. 


\bibitem{power} E.A. Power and T. Thirunamachandran, Phys. Rev. A {\bf 25}, 2473 (1982).

\bibitem{pumplin} J. Pumplin, Am.J.Phys. {\bf 37}, 737 (1969).

\bibitem{especiais} M. Abramowitz and I. Stegun, Eds. Handbook of Mathematical Functions, section 9.6. Dover (1972).

\bibitem{griffiths} D.J. Griffiths, Introduction to Electrodynamics, Prentice Hall (1999).


\bibitem{fila} F.C. Santos and A.C. Tort, Eur.J.Phys., {\bf 25}, 859.

\bibitem{taddei} M.M. Taddei, T.C.N. Mendes and C. Farina, arXiv:0903.2091. For a more simplified demonstration
of the atom-sphere dispersive force see Eur.J.Phys. {\bf 31}, 89 (2010).

\bibitem{Thiru} D.P. Craig and T. Thirunamachandran, pg. 158-159, Molecular quantum electrodynamics. New York: Dover. (1998)




%
%
%
%
\end{thebibliography}
\end{document}